%Paper: astro-ph/9503079
%From: Edvige Corbelli <edvige@arcetri.astro.it>
%Date: Tue, 21 Mar 1995 14:51:37 +0100

\magnification=\magstep1
\baselineskip=16pt
\vsize=8.9 true in
\hsize=6.5 true in
\hoffset=-0.1truecm\voffset=0.3truecm
\nopagenumbers\parindent=0pt
\footline={\ifnum\pageno<1 \hss\thinspace\hss
    \else\hss\folio\hss \fi}
\pageno=-1

\newdimen\windowhsize \windowhsize=13.1truecm
\newdimen\windowvsize \windowvsize=6.6truecm

\def\simless{\mathbin{\lower 3pt\hbox
   {$\rlap{\raise 5pt\hbox{$\char'074$}}\mathchar"7218$}}} %< or of order
\def\simgreat{\mathbin{\lower 3pt\hbox
   {$\rlap{\raise 5pt\hbox{$\char'076$}}\mathchar"7218$}}} %> or of order
\def\ltsima{$\; \buildrel < \over \sim \;$}
\def\simlt{\lower.5ex\hbox{\ltsima}}
\def\gtsima{$\; \buildrel > \over \sim \;$}
\def\simgt{\lower.5ex\hbox{\gtsima}}
\def\ref{\noindent\hangindent.5in\hangafter=1}
\def\heading#1{
    \vskip0pt plus6\baselineskip\penalty-250\vskip0pt plus-6\baselineskip
    \vskip2\baselineskip\vskip 0pt plus 3pt minus 3pt
    \centerline{\bf#1}
    \global\count11=0\nobreak\vskip\baselineskip}
\count10=0
\def\section#1{
    \vskip0pt plus6\baselineskip\penalty-250\vskip0pt plus-6\baselineskip
    \vskip2\baselineskip plus 3pt minus 3pt
    \global\advance\count10 by 1
    \centerline{\expandafter{\number\count10}.\ \bf{#1}}
    \global\count11=0\nobreak\vskip\baselineskip}
\def\subsection#1{
    \vskip0pt plus3\baselineskip\penalty-200\vskip0pt plus-3\baselineskip
    \vskip1\baselineskip plus 3pt minus 3pt
    \global\advance\count11 by 1
    \centerline{{\it {\number\count10}.{\number\count11}\/})\ \it #1}}
\def\firstsubsection#1{
    \vskip0pt plus3\baselineskip\penalty-200\vskip0pt plus-3\baselineskip
    \vskip 0pt plus 3pt minus 3pt
    \global\advance\count11 by 1
    \centerline{{\it {\number\count10}.{\number\count11}\/})\ \it #1}}

\def\eol{\hfil\break}
\def\affl#1{\noindent\llap{$^{#1}$}}
\def\simlt{\lower.5ex\hbox{$\; \buildrel < \over \sim \;$}}
\def\simgt{\lower.5ex\hbox{$\; \buildrel > \over \sim \;$}}

%TITOLO
{
\def\cl#1{\hbox to \windowhsize{\hfill#1\hfill}}
\hbox to\hsize{\hfill\hbox{\vbox to\windowvsize{\vfill
%%%%%%%%%%%%%%%%%%%%%%%%%%%%%%%%%%%%%%%%%%%%%%%%%%%%%%%%%%%%%%%%%
%Sostituire Titolo, Autori (con n. affiliazioni) e n. Preprint  %
%%%%%%%%%%%%%%%%%%%%%%%%%%%%%%%%%%%%%%%%%%%%%%%%%%%%%%%%%%%%%%%%%
\bf
\cl{THE ONSET OF THE COLD HI PHASE}
\cl{IN DISKS OF PROTOGALAXIES}
\bigskip
\cl{Edvige~Corbelli$^{1}$ and Edwin~E.~Salpeter$^{2}$}
\bigskip\rm
\cl{Preprint n.~9/95}

\vfill}}\hfill}}

%AFFILIAZIONI
\vskip5truecm
{\leftskip1.7truecm
%%%%%%%%%%%%%%%%%%%%%%%%%%%%%%%%%%%%%%%%%%%%%%%%%%%%%%%%%%%%%%%%%
%Sostituire n. e titolo affiliazioni				%
%%%%%%%%%%%%%%%%%%%%%%%%%%%%%%%%%%%%%%%%%%%%%%%%%%%%%%%%%%%%%%%%%
\affl{1}Osservatorio Astrofisico di Arcetri,
\eol
Largo E.~Fermi 5, I-50125 Firenze (Italy)
\bigskip
\affl{2}Center for Radiophysics and Space Research,
\eol
Cornell University, Ithaca NY, 14853 (USA)

%GIORNALE DI PUBBLICAZIONE
\vfill
%%%%%%%%%%%%%%%%%%%%%%%%%%%%%%%%%%%%%%%%%%%%%%%%%%%%%%%%%%%%%%%%%
%Sostituire nome della rivista					%
%%%%%%%%%%%%%%%%%%%%%%%%%%%%%%%%%%%%%%%%%%%%%%%%%%%%%%%%%%%%%%%%%
To appear in The Astrophysical Journal .
\vglue3truecm
}
\eject

%ABSTRACT
\vglue 5 true cm
\heading{ABSTRACT}

\vglue 0.2 true in

We discuss a possible delay experienced by protogalaxies with low column
density of gas in forming stars over large scales. After the hydrogen has
recombined, as the external ionizing
UV flux decreases and the metal abundance $Z$ increases, the HI, initially
in the warm phase ($T\simgt 5000$ K), makes a transition to the cool phase
($T\simlt 100$ K). The minimum abundance $Z_{min}$ for which this phase
transition takes place in a small fraction of the Hubble time decreases
rapidly with increasing gas column density. Therefore in the ``anemic''
disk galaxies, where $N_{HI}$ is up to ten times smaller than for normal
large spirals, the onset of the cool HI phase is delayed. The onset
of gravitational instability is also delayed, since these objects
are more likely to be
gravitationally stable in the warm phase than progenitors of today's
large spiral galaxies. The first substantial burst of star formation may occur
only as late as at redshifts $z \sim 0.5$ and give a temporary high
peak luminosity, which may be related to the ``faint blue objects".
Galaxy disks of lower column density tend to have lower escape velocities
and a starburst/galactic fountain instability which decreases the gas
content of the inner disk drastically.

\medskip
\noindent\underbar{\strut Subject}\ \underbar{\strut headings}:
galaxies: evolution - galaxies: formation -
galaxies: ISM - stars: formation

\vglue\windowvsize plus 1fill minus \vsize
\eject

%TEXT
\pageno=1
\section{ INTRODUCTION}

\vglue 0.1 true in

The complexity of the interstellar medium in the inner disk of
present-day spiral galaxies (e.g. Kulkarni $\&$ Heiles 1988) is in part
due to the formation
of massive stars: heavy elements in the gas phase and dust have been
produced by previous star formation, while present-day massive stars provide
a copious input of bulk kinetic energy and of ionizing photons. As a
consequence, heating and ionization by the extragalactic photon flux is
usually considered unimportant, except in the outermost disks where star
formation is absent and where the gas has a low column density and is
partially ionized (Maloney 1993; Corbelli $\&$ Salpeter 1993a,b;
Dove $\&$ Shull 1994). In the past, however,
before the first substantial burst of star formation occurred, most of the
material in protogalaxies was still mainly gaseous and the extragalactic
background flux played a very important role in the time
evolution and in the transition toward a starburst phase.

The radiative cooling in the earliest phases of the protogalaxy formation
was fast for objects with a relatively small virial theorem velocity
dispersion (e.g. Rees $\&$ Ostriker 1977; Silk 1977). After this initial
transient and before extensive star formation starts, the quasi-steady
state ionization and thermal balance depends on the extragalactic ionizing
flux. Proto-galaxies with small column densities of gas  (e.g. dwarf
galaxies) may experience a delay in the recombination of ionized hydrogen
until this flux has dropped below a certain value (Babul $\&$ Rees 1992;
Efstathiou 1992). In the present paper we consider the next evolutionary
phase for disk galaxies, where the circular rotation velocity $V_r$ accounts
for most of the virial velocity dispersion and remains constant.
We shall estimate the further
cooling (during this phase and after the hydrogen has recombined)
assuming that the changing sound speed $v_s$
satisfies $v_s \ll V_r$. This ensures that we can use a slab approximation
with gas column density unchanged during the cooling process
(although the scale height decreases). Evidence for rotationally
supported disks at redshifts $z\sim 2-3$ is given for example by
Wolfe {\it et al.} (1994).

For a volume of predominantly neutral hydrogen gas with solar metal abundance,
a given thermal pressure $P$ and a given ionization rate $\xi$ per neutral
H-atom, the equilibrium gas temperature depends on the ratio $P/\xi$.
Field, Goldsmith, $\&$ Habing (1969) showed that over a narrow range of values
of $P/\xi$ two phases of HI can coexist: a warm phase at
$T\sim 10^4$ K and a cool phase at $T\sim 100 $ K. A third phase found at
intermediate temperatures ($T\sim 3000$ K) is  thermally unstable. For larger
(smaller) values of $P/\xi$ only the cold (warm) phase exist. For a given
$P/\xi$ the equilibrium temperature depends also on the abundance $Z$ of
heavy elements (throughout this paper given in units of solar abundances)
and there would
not be a range of $P/\xi$ with coexisting phases if $Z=0$. We shall
consider medium-small values of $Z$, e.g. a slow build-up of metals
with time during this phase, from $Z\sim  10^{-4}$
to $Z\sim 0.1$. This assumption of slow, sporadic metal injection
is discussed in Section 2.
We will analyze in detail the possibility that after the
hydrogen has recombined, some proto-disks experience a warm to cool phase
transition before violent star formation turns on
over large scales. At which redshift this phase transition occurs
depends quite sensitively on the metallicity $Z$, on the intensity
and evolution of the extragalactic background and on the neutral
hydrogen column density $N_{HI}$. Since in
ordinary disk galaxies $N_{HI}$ varies radially and with galaxy type, we
shall consider slabs with gas distribution whose maximum gas
column density before star formation takes place,
$N_{max}$, varies from $\sim 10^{22.5}$ cm$^{-2}$, as for ordinary
large spiral galaxies, to values ten times smaller $\sim 10^{21.5}$ cm$^{-2}$.
The latter is of interest for extreme dwarf irregular galaxies and for
putative ``Cheshire Cat'' galaxies (see Section 4 and Salpeter 1993)
where $N_{max}$ is small
but the radial scale-length for $N_{HI}(R)$ may not be small.

The external radiation field (UV and X-ray background at a given redshift)
and the structure of the protogalactic disk are discussed in Section 2.
The main numerical results for the cooling from the warm HI phase to the
cool phase as function of the background flux intensity, $z$,
$Z$ and $N_{HI}$ are given in Section 3.
We shall see that results depend weakly on the dark matter
column density, on the gas radial scalelength and on the rotation velocity
$V_r$ except for cases with small values of both $N_{max}$ and $V_r$.

The motivation for studying the transition of a proto-disk from the warm to
the cold HI phase is two fold: $(i)$ temperature measurements are becoming
available for the HI in damped Ly$\alpha$ absorbers at intermediate
redshifts (e.g. $z=0.69$, Cohen {\it et al.} 1994). These systems are
likely to be proto-disks of some kinds of galaxies and it is of interest
to know whether finding warm HI must imply the presence of young massive
stars or whether it might merely be a proto-disk before the phase transition.
$(ii)$ Observations of faint blue galaxies at intermediate redshifts ($z
\sim 0.5$), but not at low $z$ have led to speculations  that a
catastrophic first starburst may have occurred with a consequent  galactic
wind and substantial mass loss. In Section 4 we present speculations
on an accelerating starburst cycle which can lead to such mass loss.
We shall show that such phenomena are likely to occur in disks with
small column densities of gas, but not in proto-galaxies which are the
progenitors of today's large spiral galaxies.

\vglue 0.1 true cm
\section
{PROTOGALACTIC DISKS AND THE EXTERNAL RADIATION FIELD}
\vglue 0.1 true cm
{\centerline {\it 2.1-The extragalactic radiation field.}}
\vglue 0.1 true cm

The UV and X-ray extragalactic background at a given redshift $z_{obs}$
depends on the sources of UV and X photons at $z>z_{obs}$ and on the
number of intervening systems like Ly$\alpha$ clouds and Lyman Limit
systems (see Section 4). The overall space distribution of quasars and
their luminosity function and contribution to the UV and X-ray
background are still subject of controversy (e.g. Bechtold {\it et al.}
1987; O'Brien, Gondhalekar, $\&$ Wilson 1988;  Hartwick $\&$ Schade 1990).
We shall be interested only in redshifts $z\simlt 2.5$,
where quasars account for either an appreciable fraction or most of the
ionizing flux (Irwin, McMahon, $\&$ Hazard 1991; Warren, Hewett, $\&$
Osmer 1991).
The proximity effect at $z\sim 2$ gives a flux value at the Lyman edge
$J_{21}\simgt 0.3\times 10^{-21}$ ergs s$^{-1}$ cm$^{-2}$ sr$^{-1}$
Hz$^{-1}$ but we are interested in higher photon energies ($E>$100 eV)
for penetrating and heating protogalactic
clouds with HI column densities $\simgt 2\times 10^{20}$ cm$^{-2}$.
In this spectral region there are
still uncertainties regarding both the intrinsic emission spectrum and the
attenuation by intervening clouds (Madau 1992; Miralda-Escude
$\&$ Ostriker 1992; Meiksin $\&$ Madau 1993).

Since photon energies of interest are above 54 eV, the resulting
background spectrum after absorption by intervening systems depends
mainly on the HeII abundance in these systems:
the total intervening average neutral hydrogen column
density of a few times $10^{19}$ cm$^{-2}$ comes mainly from the
Lyman limit systems which have individual column densities
$> 10^{17.2}$ cm$^{-2}$ and are overwhelmingly neutral. The overall
He/H abundance ratio by number is only $\sim 0.1$, the total intervening
He column density is therefore only a few times $10^{18}$ cm$^{-2}$
and the HeII
column density is even less. However, the uncertainties are much greater
for the intervening Ly$\alpha$ forest systems (which individually
have $N_{HI}<10^{17.2}$ cm$^{-2}$) because the hydrogen is mainly ionized
but the value of the large ratio $N_{HII}/N_{HI}$ depends of the physical
model for these clouds in addition to the original, unabsorbed background
spectrum (Miralda-Escude $\&$ Ostriker 1992; Giallongo $\&$ Petitjean 1994;
Madau $\&$ Meskin 1994). For example, for models with very low external
pressure and
little dark matter, the ratio $N_{HII}/N_{HI}$ can exceed $10^4$, He is
fully ionized and $N_{HeII}/N_{HI}\sim 10^3$. In those cases
the helium opacity due to Lyman-$\alpha$ forest can be large and important
for the resulting shape of the background spectrum. On the other hand
models which favor a high pressure for the forest clouds with
$N_{HII}/N_{HI}<10^3$, must have $N_{HeII}/N_{HI}<100$ and a small
attenuation of the background above 54 eV.

Given the uncertainties in the attenuation of the background spectrum
we shall introduce an adjustable multiplying factor $J_0$ for the
intensity of the background flux. As reviewed by Hartwick $\&$ Schade
(1990) and by Boyle, Shanks, $\&$ Peterson (1988)
the quasar flux decreases with
time for redshift $z\le 1.5$ and we shall use the following simple formula
for energies above 100 eV, which approximates the
spectral shape after absorption by HeII (Madau 1992) and is
consistent with a decrease of the spectral index as we approach
the hard X-rays spectrum ($\alpha=0.4$ for $E\ge 1.5$ keV):

$$J(\nu,z)=3\times 10^{-25} J_0 \Bigl({\nu \over 100~{\hbox {eV}}}
\Bigr)^{-1} (1+\bar z)^3\ {\hbox{ergs}}\ {\hbox{s}}^{-1}
{\hbox{cm}}^{-2} {\hbox{sr}}^{-1} {\hbox{Hz}}^{-1} \eqno (2.1)$$

\noindent
with $\bar z$ = 1.5 for $z> 1.5$ and $\bar z$
=$z$ otherwise. For $J_0\simeq 1.2$ the above expression
is close to Madau's estimate above 100 eV but
given the uncertainties mentioned above this value of $J_0$
might underestimate the real flux and the range of $J_0$ we shall
consider is $1\leq J_0 \leq 5$.

Another important flux of interest is the UV flux for ionizing
carbon (11.26 eV) since CII (158$\mu$m) is the most important cooling
transition in the cool HI phase. The quasar
contribution to these energies is slightly higher than at the Lyman edge
and a further contribution can derive from star-forming galaxies. We
shall show that even the lower limit to the
UV flux is already sufficient to keep the carbon singly ionized.

\vglue 0.1 true cm
{\centerline {\it 2.2-Geometry for protogalactic disks.}}
\vglue 0.1 true cm

We consider only objects which evolve into protogalactic disks with circular
rotational velocity in the range 30 km s$^{-1} < V_r < 300$ km s$^{-1}$
and with nucleon column density $N_g$ (normal column density of gas
above and below the plane, by number, hydrogen plus four
times that of helium) whose maximum values $N_{max}$ [$N_g(R=0)$] is
$10^{21.5} \simlt N_{max} \simlt 10^{22.5}$ cm$^{-2}$ before large
scale star formation starts. Giant spiral
galaxies occupy the upper end of both ranges, extreme dwarf irregulars
the bottom end, and the ranges also cover the putative ``Cheshire Cat''
galaxies (Hoffman {\it et al.} 1993; Salpeter $\&$ Hoffman 1995) before
starburst induced galactic winds decrease $N_{max}$ further (see Section 4).
As discussed before (Gott $\&$ Thuan 1976;  Rees $\&$ Ostriker 1977;
Dekel $\&$ Silk 1986; Chiba $\&$ Nath 1994)
radiative H and He cooling can lead to a partially ionized, but
predominantly neutral, gas with temperature $\sim$ 1 to 3
$\times 10^4$ K in less than a Hubble time $t_H$, at redshift $z\sim
2.5 $. The isothermal sound speed is thus at most $v_s\equiv\sqrt{k T
/\mu m_H}\sim$ (7 to 15) km s$^{-1}$ ($k$ is the Boltzman constant)
and we have the inequality
$v_s\ll V_r$ (only barely at the lower end of the ranges).
This inequality ensures that the vertical scaleheight $x_{1/2}$ of the
galactic disk is small compared with the galaxy radial scalelength.
The protogalaxy
is therefore rotationally supported against gravitational collapse.
Hydrostatic equilibrium at each radius $R$ is possible due to thermal
pressure support against gravitational attraction.
We assume the above inequality throughout this paper, so that the total
column density of gas
and $V_r(R)$ do not change with time, even though the temperature
$T(R)$ and  $x_{1/2}(R)$ decrease with time as the gas cools further.

We define the gas vertical scale height $x_{1/2}$ as the height
above the midplane such that half of the total gas mass per unit
area lies between $+x_{1/2}$ and $-x_{1/2}$.
To estimate the total gas pressure at $x_{1/2}$  we can write
the approximate simple formula assuming an almost spherical halo
of dark matter:

$${P_{1/2}(R)\over k} \simeq {P_{ext}\over k} + {\pi G m_H^2 \over
2.6 k} N_g^2(R) \Bigl\lbrace 1+\eta(R) \Bigr\rbrace {\hbox {cm}}^{-3}
{\hbox {K}} \eqno(2.2)$$

$$\eta(R)\equiv {N_{dm}(R)\over N_g(R)}{v_{s}(R)\over V_r(R)}
\eqno(2.3)$$

\noindent
where $G$ is the gravitational constant,
$N_g$ is the total column density of nucleons and $N_{dm}$ is
the column density of dark matter, defined as the total dark matter mass
per unit area divided by the proton mass $m_H$.

We assume a roughly exponential form of $N_g(R)$ in the inner disk,
so that much of the total mass is contributed by regions where
$N_g\sim e^{-2} N_{max}$ to $N_{max}$. We disregard the very innermost
scalelength because of possible complications from a nuclear bulge or
AGN activity and consider only radii where $N_g\sim (0.1$ to $0.3)N_{max}$,
so that $N_g>10^{20.5}$ cm$^{-2}$ throughout. The pure gas self gravity
term in equation [2.2] thus exceeds 200 cm$^{-3}$ K whereas $P_{ext}/k
\simlt 15$ cm$^{-3} K$ for $z\simlt 2.5$ (e.g. Charlton, Salpeter,
$\&$ Linder 1994). We therefore omit $P_{ext}$ entirely in eq.  [2.2].
Some local external pressure might not be
completely negligible if ram pressure is present with a
consequent infall of hot material which confines the disk;
for these hypothetical cases the effect on $P_{1/2}$ to a first approximation
can be considered by choosing a slightly higher value of $\eta$.
For the inner disk of
regular, bright spiral galaxies the ratio $N_{dm}/N_g \simlt 1$ (Persic
$\&$ Salucci 1990) and $V_r > 100$ km s$^{-1} > v_s$, so that the
second term in bracket in eq. [2.2] can be omitted entirely. However
towards the most extreme dwarf irregular galaxies $N_{dm}/N_g$ increases
(roughly as $L^{-0.7}$) and $V_r$ decreases so that $\eta$ in eq. [2.3],
becomes appreciable. In most of this paper we consider $(1+\eta)$ as
a constant, so that $P_{1/2}$ appears proportional to
$N_g^2$ and independent of temperature $T$. The possible
temperature-dependent pressure enhancement for extreme dwarfs is briefly
discussed in Section 3.3.

\vglue 0.1 true cm
{\centerline {\it 2.3-The gravitational instability.}}
\vglue 0.1 true cm

For a rotating disk with thickness less than the radial scalelength, the
criterion for gravitational stability does not depend strongly on the
thickness (Goldreich $\&$ Lynden-Bell 1965b). We therefore consider
only the ``Toomre criterion'' for an infinitely thin disk and arbitrary
rotation law $V_r(R)$. The condition for the gravitational instability
depends on the gas velocity dispersion which, in the absence of random
motion and for an isothermal slab, equals the sound speed $v_s$ defined
in the previous subsection.
%{\bf Ed, from what I
%understood spending the whole afternoon on the argument is the pressure
%derivative respect to column density which enters into the formula, and
%therefore the one dimensional radial velocity dispersion;
%the only doubt I have is that Cowie quotes $v_s=15$ km/s in the solar neig.
%%%which seems to me a little high. For a finite thickness we have 2
%instead of $\pi$ in the following formula}.
The condition for instability can be written as (Toomre 1964;
Goldreich $\&$ Lynden-Bell 1965a; Binney $\&$ Tremaine 1987):

$$N_{g}(R)\simgt N_g^{crit}(R)\equiv
{\kappa(R) v_s(R)\over \pi m_H G} \eqno (2.4)$$

\noindent
where the epicyclic frequency $\kappa$ is

$$\kappa(R)\equiv 1.41{V_r(R)\over R}\Bigl\lbrack 1+{R\over V_r(R)}
{dV_r\over dR}\Bigr\rbrack^{1/2} \eqno (2.5)$$

\noindent
If the dark matter has an isothermal distribution, $N_{dm}(R)\propto
R^{-1}$, the rotation law is given by

$$V_r^2(R)= \pi m_H G R \Bigl\lbrack
{4\over \pi} N_{dm}(R) + \bar N_g \Bigr\rbrack \eqno (2.6)$$

\noindent
where $\bar N_g$ is the mean gas column density inside a radius $R$.
Since we are mostly interested in $R\sim$ (1 or 2) radial scalelength
$R_l$, where $V_r(R)$
varies slowly, the bracket in eq. [2.5] can be replaced by unity and
the inequality [2.4] can then be rewritten in a dimensionless form

$${N_g^{crit}(R)\over N_g(R)} \equiv  1.41 {(4/\pi)N_{dm}(R)+
\bar N_g \over N_g(R)}{v_s(R)\over V_r(R)} \simlt 1 \eqno (2.7)$$

Instability first sets in when $N_g^{crit}/N_g$ is less than unity.
In regular spiral galaxies
at  $R\sim$ (1 or 2)$R_l$, $N_{dm}$ is not much larger than $N_g$
(Persic $\&$ Salucci 1990), $1.4v_s<30$ km s$^{-1} \ll V_r$,
and the the instability criterion
is satisfied as soon as the transient radiative cooling is over and the
hydrogen has at least partially recombined. At some larger radius $N_g(R)$
drops below the critical value for instability (Kennicutt 1989), since
$N_g\propto
e^{-R/R_l}$ whereas  $N_{dm}$ decreases only as $R^{-1}$, but in this
paper we consider only the inner disk. However, for dwarf irregulars
and for the putative ``Cheshire Cat'' galaxies $V_r$, $N_g$, $N_{dm}$, and
$N_g(R_l)/N_{dm}(R_l)$ are all smaller although it is not clear what
$N_g(R_l)/N_{dm}(R_l)$ is. For either or both these classes of galaxies
we may have the following situation: when the hydrogen first
recombines but is still in the warm phase, $v_s\sim 7-15$ km s$^{-1}$
and $N_g(R_l)$ may be below the critical value for instability,
but in the cool phase
$v_s\simlt 1$ km s$^{-1}$ and the instability condition in eq. [2.7]
may be satisfied. Thus low surface brightness objects which are
less opaque to the background radiation and have a lower rotational
speed become gravitationally unstable and collapse later, only when
the slab makes a transition to the cold phase. As a result the
star formation epoch in dwarf galaxies is delayed and, as we shall see,
it may become much more violent than in normal spiral galaxies.

\vglue 0.1 true cm
{\centerline {\it 2.4-The metal abundance.}}
\vglue 0.1 true cm

This paper will be concerned directly with the onset of a systematic
gravitational instability in the disk  and of a massive starburst plus
galactic fountain activity, producing metals. However, we have to make some
assumptions about a possible slow and minor buildup  before this
bulk star formation starts suddenly.
The presence at high redshifts of quasars and AGN is likely to have given
some metal contamination early. The extended metal line absorption
systems (which are likely to be associated with galaxies, see Section 4)
suggest in fact that some metals appeared in a nuclear bulge and halo
before any star formation in the proto-galaxy disk.  We shall start
our disk calculations
with an initial $Z \sim 10^{-4}$, rather than metal-free gas. Early
metal contamination is not likely to have been more than $10^{-3}$, as
seen from the small number of present-day halo stars with $Z \simless
10^{-3}$ (Spite $\&$ Spite 1992; Norris, Peterson $\&$ Beers 1993).

Most galaxies have smaller companions, and minor gravitational interactions
could move parcels of gas to increase gas column density in a small fraction
of the disk, resulting in some very localized star formation.
Moreover  present-day dwarf irregular galaxies have typical abundances
of $Z \simless 0.1$, but  very few star-less dwarf
proto-galaxies are known today, even though they would be easy to detect
in HI emission  (e.g. Briggs 1990; Hoffman, Lu, $\&$ Salpeter 1992). This
suggests that very small bursts of star formation (processing,
say, $10^{-4}$ or $10^{-3}$ of the gas into stars) were fairly common.
We therefore make the following assumptions:  our
explicit calculations for a gas slab are for constant column density
(no explicit external interference), but we allow the metal abundance
$Z$ to increase slowly with time (to mimic the sporadic mini-starbursts).

We assume a helium abundance by number of 0.1, so that the total column
density of nucleons is 1.4$N_H$. The metal species considered are carbon,
iron, oxygen, nitrogen, silicon, and sulfur with abundance ratios  $Z$
relative to solar. We have assumed that the ionized to neutral ratio
for oxygen and nitrogen are the same as for hydrogen to account for charge
exchange effects. There is a sufficient number of photons below 13.6 eV to
keep carbon, iron, silicon and sulphur singly ionized. For example the
condition for carbon to be all singly ionized is:

$$ {P\over k} \ll 1.3 {\xi_C \over 10^{-12}} {100\times T^{1.8}
\over Z} \eqno (2.8)$$

\noindent
the contribution to the carbon ionization rate, $\xi_C$, from the
quasars alone at $z\sim 2$ and
at 11.6 eV gives $\xi_C \sim J_0\times 10^{-12}$ s$^{-1}$ and hence
the above condition at $T=100$ K for example, is satisfied as soon
as $P/k \ll 5\times 10^5 J_0/Z $ which we shall see is always true.

Since in this paper we study the evolution of proto-disks before a large
scale star formation takes place and assume a relatively low values of $Z$,
we neglect dust grains and their possible contribution to gas heating.

\vglue 0.1 true cm
\section {THE TRANSITION FROM THE WARM TO COOL HI PHASE }
\vglue 0.1 true cm

We are interested in the structure of a slab, as a portion of a
protogalactic disk, with a fixed value of the gas column density $N_g$.
The vertical equilibrium of the slab changes with time because the incident
ionizing flux $J_0(1+\bar z)^3$ in eq. [2.1] decreases for redshifts
lower than $z=1.5$ and the metal
abundance $Z$ is assumed to increase with time. Although we do not
specify the rate of increase in $Z$ (due to sporadic star formation
elsewhere), changes in flux and in $Z$ decrease the heating and
increase the cooling, so that the equilibrium temperature
of any gas parcel decreases with time. The gas stays
warm and partially ionized until a certain epoch which depends on
$N_g$ both through the pressure term and through the opacity which
the slab offers to the external UV radiation.
We consider only values of $N_g\simgt 10^{20.5}$ cm$^{-2}$ and
$J_0(1+\bar z)^3\simlt 200$, in which case the ionized layer at large heights
$x$ contains little mass. For given values of
$N_g$, $z$ and $Z$, the pressure $P(x)$ increases with decreasing
height $x$ and the surviving ionization rate $\xi(x)$ decreases because
of absorption by matter above $x$. The temperature therefore decreases
with decreasing $x$ and also with increasing time. Because cooling
rates are finite, there is a time-delay between temperature actually
achieved and the thermal balance equilibrium temperature. This delay is
discussed in Section 3.2; we first calculate equilibrium temperatures
for fixed values of $Z$ and $z$.

\vglue 0.1 true cm
{\centerline {\it 3.1-The transition point for the two-phase equilibrium.}}
\vglue 0.1 true cm

At first (large $z$, small $Z$) most of the
hydrogen is in the warm HI phase.
In the absence of any metals there is just a single phase medium and
the gas temperature hardly gets much below 5000 K, but with even a
small amount of metals a two phase medium becomes possible.
We shall call ``transition point'' the lowest
equilibrium temperature for which the warm phase is thermally stable.
In principle, for a slab of uniform composition the sharp transition
between the warm and the cool phase starts in the midplane ($x=0$)
and moves up as $Z$ increases and $z$ decreases. In this paper
we shall consider only the evolution
of a single characteristic parcel of gas at height $x_{1/2}$.
The total gas pressure for this parcel of gas, $P_{1/2}$ as given by
eq. [2.2] is a good approximation in the
warm phase since the gas is quite uniform in temperature. Our
numerical calculations show, moreover, that the transition
between a uniform warm slab and a cold slab,
where most of the gas is at temperature 30 K, is quite sharp:
once a cold core forms the warm atmosphere which is
left above is small and therefore $P_{1/2}$ is again a good
approximation. For an accurate time dependent calculation we have the
complication that $\eta$ decreases with time as $\sqrt{T}$ but
in this subsection we consider $(1+\eta)$ as a constant factor.

Background photons are responsible for the ionization-recombination
balance of the hydrogen and helium gas. Secondary electrons
are included as well and the on the spot approximation is used.
Close to the transition point the hydrogen gas is mostly neutral,
and therefore for the photon energies which are of interest, helium
is singly ionized with fractional ionization 3 times that of hydrogen.
We keep track of corrections for ionization in
an explicit calculation but $N_{HI}$ is close to the total hydrogen
column density, and the photon absorption, which depends strongly on
$N_{HI}$, changes little with time.
Heating due to photoionization of H and He by background radiation
is balanced by radiative cooling via CII, SiII, FeII, OII, OI, NI, HI lines.
At very low temperatures ($T<50$ K) additional heating comes from carbon
photoionization. The expression for energy losses due to collisional
excitation of line radiation have been taken from Dalgarno $\&$ McCray (1972)
except for the fine structure excitation of CII and OI by atomic hydrogen
impact, for which we have used the results of Launay $\&$ Roueff (1977),
and for the Ly-$\alpha$ excitation of neutral hydrogen for which we refer to
Spitzer (1978).

In Figure 1 we show two examples of the curve which determines the equilibrium
between the phases (Field {\it et al.} 1969). In principle the equilibrium
temperature $T$ is a function of the ratio of $P_{1/2}$ to photoionization
rate per atom $\xi_{1/2}$, but since we have fixed
the HI column density in order to compute the absorption of the UV flux,
the ratio $P_{1/2}/(1+\eta)$ of each slab is fixed and $\xi_{1/2}$
for a fixed spectral index $\alpha$ depends linearly on the
intensity $J_0(1+\bar z)^3$ of the incoming radiation.
For $N_{HI}=3\times
10^{20}$ and $N_{HI}=3\times 10^{21}$ cm$^{-2}$ the ratio $P_{1/2}/(1+\eta)$
is $\sim 200$  and $\sim 20000$  cm$^{-3}$ K respectively. In Figure 1 the
vertical axis increases downward and gives $J_0(1+\bar z)^3/(1+\eta)$
required to result in the equilibrium temperature $T$, for a single value
of $Z$ and two values of $N_{HI}$. The region where
$J_0(1+\bar z)^3/(1+\eta)$ increases with increasing $T$ is thermally
unstable (Field 1965; a complete analysis of the thermal
stability in a photoionized medium is given by Corbelli $\&$ Ferrara
1995). The circle on each curve denotes the transition point, the
minimum value of temperature, $T_{tr}$, for which the stable warm phase
exists. Here the most important metal line cooling
comes from FeII. The equilibrium in the cold phase which
corresponds to the same value of $[J_0(1+\bar z)^3/(1+\eta)]_{tr}$
has instead a cooling function dominated by the CII (158$\mu$m) line
excited via H impact.

We can display the transition point (for given values of $z$ and
$J_0/(1+\eta)$) in terms of a function $Z_{tr}$ of $N_{HI}$:
$Z_{tr}$ is defined as the value of the
metal abundance $Z$ needed for the equilibrium temperature to be equal
to the transition temperature. We have carried out numerical
calculations of $Z_{tr}$ in full thermal equilibrium,
for a number of values of the
various parameters and for the assumed spectral index $\alpha=1$ in eq.
[2.1]. If one had assumed a steeper spectrum the resulting $Z_{tr}$
would depend even more strongly on $N_{HI}$.
Let $t_{heat}$ and $t_{cool}$ be the heat content divided
by the photon heating rate and radiative cooling rate, respectively, at
the transition temperature $T_{tr}\simlt 8000$ K (shown by the circles in
Figure 1 for two values of $N_{HI}$). At equilibrium $t_{heat}=t_{cool}$;
$t_{cool}$ is proportional
to the inverse of the volume density, $1/n$, and also approximately to
$1/Z_{tr}$ since metal line cooling is important below 8000 K.
Although $1/n$ decreases with increasing
$N_{HI}$, $t_{cool}$ actually increases
due to the stronger attenuation of the background flux with a consequent
decrease both of $Z_{tr}$ and of the cooling rate
at transition ($T_{tr}$ decreases with $N_{HI}$).
In Section 3.2 we discuss the value of $Z(N_{HI})$ needed in order to have
a fast warm/cool phase change.

\vglue 0.1 true cm
{\centerline {\it 3.2-Time delay for the actual transition.}}
\vglue 0.1 true cm

For values of  $J_0(1+\bar z)^3/(1+\eta)$ slightly smaller
than the transition value, both stable phases are in principle
possible. Under some circumstances it may not be obvious which phase
is actually present, but there is no controversy for the evolution we
consider here: since $Z$ (and therefore cooling) increases with time
and the ionizing flux decreases, the equilibrium temperature decreases
monotonically with time. For the actual temperature $T(t)$ to be able to
decrease with time, cooling must exceed heating slightly, i.e. the
temperature must be slightly larger than the equilibrium value. The actual
temperature $T(t)$ can never cross the equilibrium curve illustrated
in Figure 1 and will remain to the right of the warm phase portion of
the curve until the transition point is reached. We can then ask how big
the time delay is, as a function of $z$ and $N_{HI}$ in order to complete
the phase transition.
We have also investigated the time delay in recombination:
the recombination time $t_r$ decreases with increasing column density
but it stays always much smaller than $t_H$ for $N_{HI}> 10^{20.5}$
cm$^{-2}$. The Hubble time $t_H$, at redshift $z$, is defined as:

$$ t_H = 2.06\times 10^{17}{100 {\hbox {km}}{\hbox { s}}^{-1}
{\hbox {Mpc}}^{-1}\over H_0}(1+z)^{1.5}  {\hbox {s}} \eqno (3.1)$$

\noindent
where $H_0$ is the present value of the Hubble constant (we use
$H_0=75$ km s$^{-1}$ Mpc$^{-1}$ in the rest of this paper).

We compute the cooling curve at transition, keeping $J_0$, $\eta$ and
$z$ fixed, but perturbing the value of the metallicity to a value
$Z>Z_{tr}$. During the isobaric transition from the warm to
the cool phase, the volume density $n$, $x$ and $T$  evolve with time.
We call $t_{tr}$ the total time required to complete this transition from
$T=T_{tr}$ to the final temperature $T=T_f$.
For $z=1$, $J_0=1$, $\eta \ll 1$ and $N_{HI}=4\times 10^{21}$ cm$^{-2}$
we show as an example in Figure 2 the actual cooling curve given a
perturbed value of metallicity $Z=2 Z_{tr}\simeq 3.2\times 10^{-4}$.
For this case $t_{tr}$
is slightly larger than $t_H$, which in other words means that one should
have $Z \gg Z_{tr}$ in order to have a rapid cooling to the cool phase.
Figure 3$(a)$
gives the ratio of $t_{tr}$ to the Hubble time $t_H$ as a function of
$N_{HI}$. The open circle marks the case shown in detail in Figure 2
and the star marks the column density, $N_*$, where $t_{tr}/t_H=0.2$.
We shall talk of ``rapid cooling'' when $t_{tr}/t_H\le 0.2$ and
define $Z_{min}$ for a chosen value of $N_{HI}$
as that value of $Z$ which gives $t_{tr}/t_H=0.2$.
In other words we do the following trial and error process:
we guess a value of $Z$ above $Z_{tr}$ and carry out a
time dependent heating/cooling calculation numerically, starting at a
temperature $T=T_{tr}$ keeping $Z$, $J_0$, and $z$ fixed during
the integration. Since $Z>Z_{tr}$, cooling exceeds heating already
and the imbalance increases as $T$ decreases monotonically with time.
Keeping the pressure constant during the cooling process we note
the actual time taken for $T$ to reach the cool phase, $t_{tr}$, we
adjust the value of $Z$ until $t_{tr}=0.2 t_H$  and call $Z_{min}$
this value of $Z$.
For $Z\sim Z_{tr}$ we have seen that $t_{cool}$ increases with
$N_{HI}$, therefore in order to have a fast warm/cool phase change the
metallicity should  increase by a larger factor with respect to
$Z_{tr}(N_{HI})$ as $N_{HI}$ increases. Due to the strong decrease
of $Z_{tr}$ with $N_{HI}$ the resulting $Z_{min}$ still decreases with
$N_{HI}$. In other words, if we keep constant the cooling time, the
increase of volume density with $N_{HI}$, due to self gravity, is strong
enough to require decreasing $Z_{min}$ values with increasing $N_{HI}$.
In Figure 3$(b)$ $Z_{min}(N_{HI})$ is
shown as a thick curve and $Z_{tr}(N_{HI})$ as a thin one. The ratio
$(Z_{min}-Z_{tr})/Z_{tr}$ increases with $N_{HI}$ and we mark with a
star the column density $N_*$ where $Z_{min}=2 Z_{tr}$.
While $Z_{tr}$ depends only on the ratio $J_0/(1+\eta)$, $Z_{min}$
depends on $J_0$ and $\eta$ separately since it is directly related
to the cooling time.

In Figure 4 we give the main results of this paper,
namely $Z_{min}$ as a function of $N_{HI}$ for various $J_0$, $\eta$ and
$z$. For each case we again mark with a
star the column density where $Z_{min}=2 Z_{tr}$.
Consider first values of $N_{HI}$ appreciably less
than $N_*$ at a time when $Z$ has increased to just above $Z_{tr}(N_{HI})$:
in this case a slight further excess of $Z$ above $Z_{tr}$
will ensure that the phase transition proceeds all the way to the cool
phase in a fraction of the Hubble time. By contrast, consider a case of
$N_{HI} > N_*$, e.g. $N_{HI}=6\times 10^{21}$ cm$^{-2}$ for $J_0=1$,
$\eta<<1$, and $z\sim 2.5$. Here $Z_{tr}$ would be only $10^{-5}$. But
cooling proceeds slowly for these larger column densities since $Z$ will
have to increase by at least a factor 10 above $Z_{tr}$
(to a value close to $Z_{min}$) in order to cool rapidly below 100~K.
However if, as in
this case, $Z_{min}$ is less than the likely initial abundance
$Z_{init}$, the whole cooling process happens as soon as the gas recombines.

As the redshift decreases below 1.5 different column densities will change
phase at different redshift due to the decreasing intensity of the
background flux. For example if $J_0/(1+\eta)=5$, and $Z\sim $
0.01, does not change with time between $z=1.5$ and $z=0$
all slabs with $N_{HI}$ between $5\times 10^{20}$ and
$10^{21}$ cm$^{-2}$ will undergo the phase transition.

\vglue 0.1 true cm
{\centerline {\it 3.3 - Time dependence of dark matter gravity.}}
\vglue 0.1 true cm

The dimensionless factor $\eta$ in eq. [2.2] and [2.3] represents the
contribution to gas pressure made by dark matter gravity.
As the gas cools its vertical scale height decreases, even if the dark
matter distribution is unchanged. As the height $x_{1/2}$ decreases less
dark matter resides below the layer and  $\eta$ decreases roughly as
$v_s\propto \sqrt T$ in eq. [2.3]. In an accurate time-dependent cooling
calculation one should therefore consider the time dependence of $\eta$.
In Figure 4 we show results for a large and fixed value of $\eta$,
evaluated at a fixed temperature, lower than
$T_{tr}$ and intermediate between the warm and cool phase temperatures,
say $T\sim 2000$ K. To compute approximately the transition time, we use
the prescription of Section 3.2 referring
mainly to the epoch when the warm phase of HI has gone thermally unstable
and the temperature has dropped a little below the equilibrium value at
the last stable point.

For giant spiral galaxies with large $V_r$ the factor $\eta$ is
not too large, but can be quite appreciable for dwarf irregular and
``Cheshire Cat'' galaxies, at least while the HI is in the warm phase.
Cases where $\eta$ starts large and decreases to small values as the HI
cools are of particular interest in Section 4, since the dimensionless
ratio in eq. [2.7] is of the same order of magnitude as $\eta$ and its
decrease is of interest for the gravitational instability.
Cases where
$N_g^{crit}/N_g$ starts a little larger than unity and decreases might be
gravitationally stable in the warm phase and unstable in the cool phase:
these cases will have $\eta$ not much smaller than unity  and varying
and its value cannot be neglected in the pressure equation. With $(1+\eta)$,
and hence the gas pressure $P_{1/2}$, decreasing as the gas cools, the
cooling history is more complex than the description in Section 3.2, but
not qualitatively different: the cooling rate (per H-atom) depends on gas
density $n\propto P/T$ which can be written in the form $n\propto
(T^{-1}+b T^{-1/2})$ and increases with decreasing $T$ whether the second
term dominates or not. The correct cooling rate thus increases more
slowly as $T$ decreases appreciably. While
$J_0(1+z)^3$ and $Z$ both evolve in a direction so as to accelerate the
cooling transition, the decrease in $(1+\eta)$ decreases the pressure and
hence tends to slow down the cooling somewhat.

\vglue 0.7 true cm
{\bf{4.A POSSIBLE STARBURST/GALACTIC FOUNTAIN INSTABILITY:}}

\noindent
{RELATIONS TO QSO ABSORPTION LINES AND TO FAINT BLUE GALAXIES}
\vglue 0.7 true cm

This paper deals directly only with the inner gaseous disks of
protogalaxies, but their evolution is related  to the faint blue galaxies
found at intermediate redshifts, and to Ly-$\alpha$ absorption clouds
provided by the outer galaxy extensions.
Quasars were already present at very early cosmological epochs (redshifts
$z > 4$) and Ly$\alpha$ and metal absorption lines have been observed
over a broad range of redshifts and column densities.  Three ranges of
neutral hydrogen column densities are particularly well studied, the
``Ly$\alpha$ forest" with $N_{HI} \simgreat 10^{14}$ cm$^{-2}$,
the ``Lyman Limit systems" (LLS) with $N_{HI} > 10^{17}$ cm$^{-2}$ and
the ``damped wing systems" with $N_{HI} > 10^{20}$ cm$^{-2}$.  The damped
wing systems are usually associated with the inner disks of protogalaxies,
already rotationally supported at $z \sim$ 2 to 3.  At least for large
spiral protogalaxies, the star formation and metal abundances in the
inner disk at that epoch were probably at an intermediate evolutionary
stage:  In one well-documented case at $z = 2.3$, for instance (Wolfe et al.
1994), metal-production by massive stars had reached about 10\% of the
present-day Galactic value.  The LLS
(e.g. Bergeron $\&$ Boiss\'e 1991; Lanzetta $\&$ Bowen 1992; Steidel,
Dickinson, $\&$ Persson 1994a) are now generally considered to
indicate some kind of rather extended halos of disk galaxies or
proto-galaxies, with radii of order 50 kpc and fairly metal-rich,
but the high column density LLS presumably come from inner proto-disks.
The nature of the more plentyful Ly-$\alpha$ forest is still
controversial, but the metal abundance is lower (although probably
above $Z \sim 10^{-4}$) even at large redshifts (Lu 1991). Recent
observations have suggested very large sizes for high redshift ``forest
clouds", 40 to 400 kpc (e.g. Bechtold {\it et al.} 1994), and also large
extensions around ordinary galaxies at lower redshifts.  Self gravity and
temperatures $< 2 \times 10^4 K$ ensure that cloud scale-heights are
much less than the total radius of the extension (see, e.g. Charlton {\it
et al.} 1994).  Although much of the absorption at $z < 1 $ is
directly associated
with ordinary, visible galaxies, an appreciable (but controversial) fraction
comes from invisible objects (possibly 80\%; see e.g. Mo $\&$ Morris 1994).
These invisible objects can be the ``vanishing
Cheshire Cat" galaxies, discussed below.

The main postulate for protogalaxies of the ``vanishing Cheshire Cat"
type (Salpeter 1993) is that they were qualitatively similar to ordinary
disk galaxies (including large, but low-density, gas extensions), but
with a smaller maximum value $N_{max}\sim 10^{21.5}$ cm$^{-2}$
of the initial disk gas column
density, and smaller rotational velocity $V_r$.
The ratio of dark matter to gas (parametrized in Section 2)
is not specified  but is probably comparable to the ratio
for ordinary galaxies. Because of the scaling
of column densities, the gravitational escape velocity is smaller than
for an ordinary galaxy with the same radial scalelength.
Recent observations (Salpeter $\&$ Hoffman 1995)
suggest that these anemic galaxies occur in association with ordinary
galaxies as members of a loose galaxy group or as satellites of a large
ordinary galaxy.

Consider now the inner disk in three
different types of protogalaxies: $(i)$ ordinary spiral galaxies with
$V_r\sim 200$ km s$^{-1}$, an escape velocity slightly larger, and
a maximum baryon disk column density $N_{max}\sim 10^{22.5}$ cm$^{-2}$;
$(ii)$ dwarf irregular galaxies with $N_{max}$ only slightly smaller (Lo
1993) but $V_r$ much smaller (30 to 80 km s$^{-1}$)
and $(iii)$ the putative ``Cheshire Cat'', anemic galaxies.
As discussed in Section 2, we assume that a metal abundance of $Z \sim$
(1 to 3) $\times 10^{-4}$ solar was already present at a very early epoch in
all protogalaxies. We then find that ordinary spiral galaxies with large
$V_r$ were ``doubly safe'' in starting star formation early :
$(a)$ as seen in Figure 3 for large column densities
the required metal abundance for the HI to start cooling
to the cool phase is lower and the phase transition can start
at a higher redshift, even if it proceeds slowly until $Z$ increases
to about the value of $Z_{min}$ (Figure 4).
$(b)$ With dark matter not very important in the
inner disk and with $V_r>100$ km s$^{-1}$, eq.[2.7] shows that
gravitational instability could set in even if the HI were still in the
warm phase. Moreover with $V_{esc}>V_r$ and large, even a substantial
starburst is not likely to lead to a galactic wind.

The situation is different for disks with small $V_r$ and $V_{esc}$,
i.e. both dwarf irregulars and the postulated anemic protogalaxies;
especially for the latter where the central column densities are
even smaller. As seen in Figure 4, for $N_{HI}\simlt 10^{21}$ cm$^{-2}$
cooling below the warm phase at $z\sim 1$ requires $Z\simgt 0.001$
which implies some metal enrichment in the disk. Although
we don't know $N_{dm}/N_g$ in these galaxies, it is likely
that the gravitational instability (eq.[2.7]) sets in strongly only when
the transition to the cold phase, which for these galaxies happens very fast,
lowers $v_s$ by about a factor 10.
The gravitational instability triggers star formation (Kennicutt 1989;
van der Hulst {\it et al.} 1993; Taylor {\it et al.} 1994; Dopita
$\&$ Ryder 1994) and as star formation starts in some areas, it increases
$Z$ and speeds up the transition to the cool phase for surrounding regions.
It is difficult to be more quantitative since this happens
at a later epoch, when the background flux
is  already decreasing with time and this condition favors the transition
to the cool phase and the starbursts.
But there is another effect which goes in the
opposite direction: bulk motions which are setup by
the starburst, contribute to the increasing of the overall velocity
dispersion and may counteract the decrease in the atomic thermal speed.

The concluding remark is that the postulated smaller $N_{max}$
favors the delay of the onset of extensive
star formation and the identification of the  anemic proto-disk in the
starburst phase with the ``faint blue galaxies" seen at $z \sim 0.5$.
The slightly smaller central
column density and/or smaller escape velocity than in ordinary dwarf
irregular galaxies should be sufficient to decrease the column density
of the inner disk appreciably during the starburst phase
(the vanishing of the cat body) by a galactic wind and/or fountain.
As a consequence these galaxies become fainter,
to the point that they may seem to have
faded away from $z\sim 0.5$ to now, in agreement with the decreasing number
of  ``faint blue galaxies" observed from $z \sim 0.5$ to now.
The gas mass in the outer disk, where no star formation takes
place, is unaffected by the runaway or even increased by a fountain
(i.e. the smile of the cat remains or is intensified) and originates
absorption lines in the surroundings of these today's invisible galaxies.

%There is now some direct observational evidence for
%``superbubbles'' in ordinary dwarf irregular galaxies (Marlowe {\it et al.}
%1995) which almost suggest some catastrophe.  However, the subtle
%difference between ``blow out''
%and ``blow away'' (de Young $\&$ Heckman 1994; Wang 1995) is of
%importance for the conjecture of converting an
% into a vanishing ``Cheshire Cat''
%(Salpeter $\&$ Hoffman 1995):

\vglue 0.1 true cm
\section{SUMMARY AND DISCUSSION}
\vglue 0.1 true cm

We have discussed an intermediate phase in the evolution of
protogalaxies which takes places after the hydrogen has recombined
and a circular disk has formed but before the first substantial burst
of star formation. This phase is regulated by the slow decrease in time of
the extragalactic ionizing photon flux and by the slow increase of the heavy
element abundance $Z$ in the ISM due to contamination from sporadic,
small-scale
star formation elsewhere in the disk (possibly due to interaction with minor
satellites). In particular we have
calculated the minimum value $Z_{min}$ which must be exceeded for the HI in
the disk to cool from the warm phase to the cool phase in a small fraction of
the Hubble time. Values of $Z_{min}$ are given in Fig. 4, as a function of the
total neutral hydrogen column density $N_{HI}$ normal to the galactic disk.
$Z_{min}$ decreases slowly with decreasing redshift $z$ and depends on the
intensity of the background at a fixed $z$ and somewhat on the dark matter
parameter $\eta$. The most important feature is however its rapid decrease
with increasing $N_{HI}$.

Disregarding the innermost part of disk galaxies, which may be affected by
nuclear activity, we focus on the inner galactic disk where the average
column density $N_{HI}$ is 0.1 to 0.3 its initial maximum central value
and where the first large scale starburst is likely to occur.
We are particularly interested in the evolutionary fate of  possible
anemic protogalactic disks (which might be either a separate class of
galaxies or merely an extension of dwarf irregulars and/or Malin objects)
in contrast with the fate of large spirals. We postulated that their
column density is slightly smaller than for irregular galaxies which in turn
is smaller than the initial column density
of today's spiral galaxies ($\sim 10^{22}$ to $10^{22.5}$ cm$^{-2}$).
Assuming $V_r < 100$ km s$^{-1}$ for ``anemics'' as it is for
dwarf irregular galaxies, the initial cooling of the ionized hydrogen
down to $\sim 3\times 10^4$ K, say, was rapid but because of the small
$N_{HI}$ and the consequent  better penetration of the extragalactic UV
flux, the H-recombination was slow (Babul $\&$ Rees 1992; Efstathiou 1992).
We have shown that these objects experience a further time delay for
cooling to temperatures below 100 K because they require the infusion of
more metals.
If one assumes that contamination gave $Z_{init}\sim 10^{-4}$ to $10^{-3}$
in all disks (Lu 1991; Spite $\&$ Spite 1992) before local star formation,
one contrast is seen immediately from Fig. 4: large spirals had $Z_{min}
< Z_{init}$ even at the largest redshifts, whereas $Z_{min} \gg Z_{init}$
even today ($z=0$) for the anemic proto-disks.
As shown by eq. [2.7] the small rotational velocity in low
column density objects may inhibit large scale star formation while the
gas is in the warm phase. Here again there is a contrast between
normal spirals with large $V_r$, where instability can set in
easily even when the sound speed is large, and dwarf irregulars or
``anemics'' for which gravitational instability might set in only
after a slow build-up of metals,
when the gas is in the cool phase and the sound speed becomes 10 times
smaller than in the warm phase.

Cohen {\it et al.} (1994) have found an HI disk at $z\approx 0.69$ with a
gas temperature $T> 1000 K$ (well above the cool phase temperature) with
a metal abundance $Z\sim (0.01$ to 0.1) and an observed column density
$N_{HI}/$cos$i=2\times10^{21}$ cm$^{-2}$ where $i$ is the disk inclination
angle. The warm temperature of the ISM can be due to star formation already
present in the disk, but we can also use Fig. 4 to ask whether - as an
alternative model - the HI may not yet have made the transition from the
warm to the cool phase. As an example take $Z=0.03$: for large values
of $J_0$, $J_0=5$, and $\eta\ll 1$ we would need
$N_{HI} < 4 \times 10^{20}$ cm$^{-2}$, to give $Z_{min}>Z$
i.e. $\cos i$ would have to be less than 0.2 for the HI disk to still
be in the warm phase, before the onset of massive star formation.
Such a small $\cos i$ may seem unlikely, but the actual value is
uncertain because of the uncertainties in $Z$.
Steidel {\it et al.} (1994b) have some optical data which argues
against the presence of many stars:  they find a very small upper limit to
optical surface brightness in the immediate vicinity of the quasar line of
sight where $N_{HI}/\cos i$ has been measured (although a LSB galaxy
is visible about 15 kpc away).

The transition to the cool phase for anemic galaxies starts later but
proceeds fast since $N_{HI}<N_*$.
When an isolated protogalaxy has the HI in its cool phase
star formation may start more easily not only because is gravitationally
more unstable but because the Jeans mass is lower, and furthermore
cool temperatures facilitate molecule formation.
In these galaxies the overall conditions are such that
the first occurrence of instability, fragmentation and formation of massive
stars may lead to a ``run-away violent starburst'', described in detail in
Section 4. The likely galactic winds or fountains decreases the gas content
of the inner disk sufficiently to shut off further star formation and the
increase of metal content. We have dubbed
extreme cases of this kind ``Cheshire Cat'' galaxies (Salpeter 1993;
Salpeter $\&$ Hoffman 1995) where the inner disk (body)  fades to almost
invisibility while the low-density outer disk (smile) remains (or is even
enhanced by an outer fountain; Bregman 1980; Corbelli $\&$ Salpeter
1988).

The less drastic version of anemic
galaxies may already explain the ``evolution puzzle'' for the faint
blue objects (e.g. Koo $\&$ Kron 1992): the evolution seems puzzling in
the context of normal, large spiral galaxies where the  onset of
star formation occurred early, $z>1$, and gave a high onset luminosity
with rapid initial fading, but slow luminosity
changes subsequently, i.e. little fading on the average from $z\sim 0.5$
to now. The faint blue objects instead seems to have experienced
a late burst of star formation and to have been more common at
modest redshifts, $z\sim 0.5$, than now. We propose that since
anemic galaxies  had their first starburst later, because of their
delay  in HI cooling and in reaching the condition for gravitational
instability, they could be identified with the
faint blue objects which then fade into very low luminosity dwarf
irregular galaxies because of the mass loss driven by the delayed
violent starburst.  To explain the optical fading alone a small mass
loss would be sufficient, but some data on high column density
Lyman Limit systems at low
redshift (Storrie-Lombardi {\it et al.} 1994) may point to more drastic
mass loss from inner disks of anemic galaxies. In fact, while at high
redshifts an appreciable fraction of LLS have $N_{HI}>2\times10^{18}$
cm$^{-2}$ and presumably come from inner proto-disks, for $z<1$ very few
LLS have $N_{HI}>2\times10^{18}$ cm$^{-2}$ (see Table 2 of Salpeter
$\&$ Hoffman 1995), which may indicate that gas content of most
anemic inner disks decreased below $\sim 2\times 10^{18}$ cm$^{-2}$.

\heading {ACKNOWLEDGMENTS}

We are grateful to J. Charlton, L. Hoffman, D. Hollenbach,
G. Field and M. Rees for
helpful discussions on the subject of this paper. This work was
supported in part by NSF grant AST91-19475 and by the Agenzia Spaziale
Italiana.

\vfill
\eject
%---------------
%No helium, program osc.f.

\def\refindent{\advance\leftskip by 24pt \parindent=-24pt}

\heading {REFERENCES}

\vglue 0.1 in

\refindent
Babul, A. $\&$ Rees, M. J. 1992, MNRAS, 255, 346\par

Bechtold, J., Weymann, R. J., Lin, Z. $\&$ Malkan, M. 1987, ApJ, 315, 180\par

Bechtold, J., Crotts, A., Duncan, R., $\&$ Fang, Y. 1994, ApJ, 437, L83\par

Bergeron, J. and Boiss\'e, P. 1991, A$\&$A, 243, 344\par

Binney, J. $\&$ Tremaine, S., 1987, in Galactic Dynamics, (Princeton
University
Press:  Princeton, N.J.), p. 310\par

Boyle B. J., Shanks, T., $\&$ Peterson, B. A. 1988, MNRAS 235, 935\par

Bregman, J. N. 1980, ApJ, 236, 577\par

Briggs, F. H. 1990, AJ, 100, 999\par

Charlton, J., Salpeter, E. E. $\&$ Linder, S. M. 1994, ApJ, 430, L29\par

Chiba, M. $\&$ Nath B. B., 1994, ApJ, 436, 618\par

Cohen, R. D., Barlow, T. A., Beaver, E. A., Junkkarinen, V. T., Lyons,
R. W.,
$\&$ Smith, H. E. 1994, ApJ, 421, 453\par

Corbelli, E. $\&$ Salpeter, E. E. 1988, ApJ, 326, 551\par

Corbelli, E. $\&$ Salpeter, E. E. 1993a, ApJ, 419, 94\par

Corbelli, E. $\&$ Salpeter, E. E. 1993b, ApJ, 419, 104\par

Corbelli, E. $\&$ Ferrara, A., 1995, ApJ, in press\par

Dalgarno, A. $\&$ McCray, R. A. 1972, ARA$\&$A, 10, 375\par

%
%De Joung, D. S. $\&$ Heckman, T. M. 1994, ApJ, 431, 598

Dekel, A. $\&$ Silk, J. 1986, ApJ, 303, 39\par

Dopita, M. A. $\&$ Ryder, S. D. 1994, ApJ, 430, 163\par

Dove, J. B. $\&$ Shull, J. M. 1994, ApJ, 423, 196\par

Efstathiou, G. 1992, MNRAS, 256, 43p\par

Field, G. B., 1965, ApJ, 142, 531\par

Field, G. B., Goldsmith, D. W. $\&$ Habing, H. J. 1969, ApJ, 155, L149\par

Giallongo, E. $\&$ Petitjean, P. 1994, ApJ, 426, L61\par

Goldreich, P. $\&$ Lynden-Bell, D. 1965a MNRAS 130, 97\par

Goldreich, P. $\&$ Lynden-Bell, D. 1965b MNRAS 130, 125\par

Gott, J. R. $\&$ Thuan, T. X. 1976, ApJ, 204, 649\par

Hartwick, F. D. A., $\&$ Schade, D. 1990, ARA$\&$A, 437\par

Hoffman, G. L., Lu, N. Y., $\&$ Salpeter, E. E. 1992, AJ, 104, 2086\par

Hoffman, G. L., Lu, N. Y., Salpeter, E. E., Farhar, B., Lamphier, C.
$\&$ Roos, T.
1993, AJ, 106, 39\par

Irwin, M. McMahon, R. G. $\&$ Hazard, C. 1991, in Space Distribution
of Quasars,
ed. D. Crampton (San Francisco: ASP)\par

Kennicutt, R. C. 1989, ApJ, 344, 685\par

Koo, D. C., $\&$ Kron, R. G. 1992, ARA$\&$A, 30, 613\par

Kulkarni, S. R., $\&$ Heiles, C. 1988, in Galactic and Extragalactic
Radio Astronomy,
ed. G. L. Verschuur $\&$ K. I. Kellerman (New York:Springer-Verlag),
p. 95.\par

Lanzetta, K. M., $\&$ Bowen, D. V. 1992, ApJ, 391, 48\par

Launay, M. $\&$ Roueff, E. 1977, A$\&$A, 56, 289\par

Lo, K. J. 1993, AJ, 106, 507\par

Lu, L., 1991, ApJ, 379, 99\par

Madau, P. 1992, ApJ, 389, L1\par

Madau, P. $\&$ Meiksin, A. 1994, ApJ, 433, L53\par

Maloney, P. 1993, ApJ, 414, 14\par

%
%Marlowe, A. T., Heckman, T. M., Wyse, R. F. G., $\&$ Schommer, R. A.
%1995, ApJ, 438, 563\par

Meiksin, A. $\&$ Madau, P. 1993, ApJ, 412, 34\par

Miralda-Escude, J., $\&$ Ostriker, J. P., 1992, ApJ, 392, 15\par

Mo, H. J. $\&$ Morris, S. L. 1994, MNRAS, 269, 52\par

Norris, J. E., Peterson, R, $\&$ Beers, T. 1993, ApJ, 415, 797\par

O'Brien, P. T. Gondhalekar, P. M.. $\&$ Wilson, R. 1988, MNRAS, 233 801\par

Persic, M. $\&$ Salucci, P. 1990, MNRAS, 245, 577\par

Rees, M. J. $\&$ Ostriker, J. P. 1977, MNRAS, 179, 541\par

Salpeter, E. E., 1993, AJ, 106, 1265\par

Salpeter, E. E., $\&$ Hoffman, G. L. 1995, ApJ, in press (March issue)\par

Silk, J. 1977, ApJ 211, 638\par

Spite, M. $\&$ Spite, F. 1992, AA, 252, 689\par

Spitzer, L. 1978, Physical Processes in The Interstellar
Medium, (Wiley:
New York)\par

Steidel, C. C., Dickinson, M., $\&$ Persson, S. E. 1994a, ApJ, 437, L75\par

Steidel, C. C., Pettini, Dickinson, M. \& Persson, S. E. 1994b, AJ, 108,
2046\par

Storrie-Lombardi, L. J., McMahon, R., Irwin, M. $\&$ Hazard, C. 1994,
ApJ, 427,
L13\par

Taylor, C. L., Brinks, E., Pogge, R. W., $\&$ Skillman, E. D.
1994, AJ, 107, 971\par

Toomre, A. 1964, ApJ 139, 1217\par

van der Hulst, J. M., Skillman, E. D., Smith, T. R., Bothun, G. D., McGaugh,
S. S., $\&$ de Blok, W. J. G. 1993, AJ, 106, 548\par

%
%Wang, B. 1995, in The Physics of the Interstellar Medium and
%Intergalactic Medium, PASP Conference Series, eds. Ferrara et al.
%in press

Warren, S. J., Hewett, P. C., $\&$ Osmer, P.S . 1991, in Space Distribution
of Quasars, ed. D. Crampton (San Francisco:ASP)\par

Wolfe, A. M., Fan, X., Tytler, D., Vogt, S. S., Keane, M. J., $\&$ Lanzetta,
K. M.
1994, ApJ, 435, L101\par

\vfill
\eject

\heading
{FIGURE CAPTIONS}
\vglue 0.1 true cm

{\bf Figure 1}. Equilibrium curves for a slab of fixed metallicity
(Z=0.01 solar) and HI column density
of gas (upper curve $N_{HI}=3\times 10^{20}$ cm$^{-2}$, lower curve
$N_{HI}=3\times 10^{21}$ cm$^{-2}$). The vertical axis increases
downward and give the required ratio of
$J_0(1+\bar z)^3/(1+\eta)$ in order to have an equilibrium temperature $T$.
The two thermally stable phase (cool and warm phase) have a  negative
slope in the graph. The open circle indicates the lowest stable
temperature for the warm phase, called ``transition point'' in the
text.

{\bf Figure 2}. Given a perturbation on the metallicity at the
transition point, $\delta Z/Z_{tr}=1$, we show for a given value
of $z$, $J_0$, and $\eta<<1$, the  time evolution of
$\delta T/T_{tr}\equiv (T-T_{tr})/T_{tr}$ from the warm to the cold phase.
Time has been normalized to the Hubble time, $t_H$.

{\bf Figure 3}. For a fixed value of $z$, $J_0$, and $\eta<<1$,
in $(a)$ we show the time required to
complete the transition between the warm and the cool phase,
as a function of $N_{HI}$,
given a perturbation on the metallicity of fixed amplitude,
$\delta Z/Z_{tr}=1$. Time has been normalized to the Hubble time, $t_H$.
The open circle indicates tha case shown in detail in Figure 2.
In $(b)$ for the same value of $z$, $J_0$, and $\eta$ we plot
$Z_{tr}$  (light curve) and $Z_{min}$ (heavy curve).
The star indicates the column density for which
$t=0.2 t_H$ in $(a)$ and $Z_{min}=2 Z_{tr}$ in $(b)$.

{\bf Figure 4}. For 3 different combination of $J_0$ and $\eta$, we
give $Z_{min}$ as function of $N_{HI}$. Continuous curves are for
$z=2.5$, dotted curves are for $z=1$ and dashed
curves are for $z=0.1$.  In each plot and for the three values of the
redshifts the star indicates the column density $N_*$ for which $Z_{min}=
2 Z_{tr}$. For $N_{HI}<N_*$ a slight excess of $Z$ above $Z_{tr}$
ensures that the transition to the cool phase proceeds rapidly.

\end